\pdfoutput=1
\documentclass{ieeeaccess}
\usepackage{cite}
\usepackage{algorithm}
\usepackage{algpseudocode}
\usepackage{algorithmicx}
\usepackage{graphicx}
\usepackage{textcomp}
\usepackage{braket}
\usepackage{caption}
\usepackage{amsmath}
\usepackage{amssymb}
\usepackage{url}

\begin{document}

\doi{}

\title{Benchmarking Swarm Optimization Algorithms for Parameter Initialization in the Quantum Approximate Optimization Algorithm}

\author{\uppercase{Shashank Sanjay Bhat}\authorrefmark{1},
\uppercase{Peiyong Wang}\authorrefmark{2},
and \uppercase{Udaya Parampalli}\authorrefmark{1}}

\address[1]{School of Computing and Information Systems, The University of Melbourne, Parkville VIC 3010, Australia}
\address[2]{CSIRO Clayton, Research Way Clayton VIC 3168, Australia}

\corresp{Corresponding author: Shashank Sanjay Bhat (email: shashankb@unimelb.edu.au).}

\begin{abstract}
The Quantum Approximate Optimization Algorithm (QAOA) is a prominent variational algorithm for solving combinatorial optimization problems such as the Max Cut problem. A key challenge in QAOA is the efficient identification of variational parameters $(\gamma, \beta)$ that yield high-quality solutions. In this work, we investigate swarm optimization methods as robust strategies for exploring the QAOA parameter space. We evaluate Particle Swarm Optimization (PSO), Fully Informed Particle Swarm Optimization (FIPSO), Quantum Particle Swarm Optimization (QPSO), and an Adam-assisted FIPSO variant on weighted MaxCut instances across multiple system sizes, circuit depths, and noise regimes, including shot noise. Our results show that these methods achieve lower approximation gaps and more stable convergence compared to standard optimizers such as Adam, COBYLA, and SPSA. In particular, we observe that swarm methods maintain superior performance under noisy and shot limited conditions. These findings suggest that population based search is effective for navigating the complex QAOA landscape and is a promising approach for parameter optimization in near-term quantum algorithms.
\end{abstract}

\begin{keywords}
Quantum Approximate Optimization Algorithm, Particle Swarm Optimization, Approximation Gap
\end{keywords}

\titlepgskip=-15pt
\maketitle

\section{Introduction}
\label{sec:introduction}
The Max Cut problem is a fundamental combinatorial optimization task in which the vertices of a graph are partitioned into two disjoint sets such that the total weight of edges crossing the partition is maximized. The problem is NP hard and arises in a broad range of applications, including statistical physics, network science, and machine learning \cite{Karp1972}. Due to its computational complexity and practical relevance, Max Cut has become a standard benchmark for evaluating optimization algorithms.

Classical approaches to Max Cut include exact methods such as branch and bound\cite{Land2010}, as well as heuristic techniques based on local search\cite{aarts2003local}. In addition, semidefinite programming relaxations, most notably the Goemans-Williamson algorithm \cite{GoemansWilliamson1995}, provide strong approximation guarantees and have demonstrated excellent empirical performance. Despite these advances, many classical methods incur significant computational overhead as the problem size increases, particularly for large and dense graphs. This limitation motivates the exploration of alternative computational paradigms. In this context, quantum algorithms have emerged as a promising direction, especially in the near term regime where hybrid quantum classical frameworks can be deployed on noisy intermediate scale quantum devices.

One such approach is the Quantum Approximate Optimization Algorithm \cite{Farhi2014}, which is well suited for combinatorial optimization problems such as Max Cut. QAOA prepares candidate solutions using a parameterized quantum circuit defined by angles $\gamma$ and $\beta$, while a classical optimizer updates these parameters to improve solution quality \cite{Egger2021}.

This framework is especially relevant in the current Noisy Intermediate Scale Quantum (NISQ) era \cite{Preskill2018}, where quantum devices are limited in size and affected by noise and decoherence. These constraints have driven the development of hybrid quantum classical methods that combine the strengths of both paradigms. In such approaches, quantum circuits perform state preparation and objective evaluation, while classical optimization routines iteratively refine the parameters \cite{McClean2016}.

Optimizing the variational parameters of QAOA remains a challenging task due to the presence of barren plateaus \cite{McClean2018} and numerous local minima in the optimization landscape \cite{Wierichs2020}. These features can hinder convergence to global minima of the QAOA landscape and render the performance highly sensitive to the choice of the initialization. As a result, effective strategies for identifying promising regions of the parameter space are essential.


In this work, we investigate swarm optimization algorithms for identifying near optimal parameters for QAOA. By leveraging population based search strategies, these methods enable efficient exploration of the QAOA parameter landscape while maintaining stable convergence behavior. In particular, we consider Particle Swarm Optimization, Quantum Particle Swarm Optimization, Fully Informed Particle Swarm Optimization, and an Adam assisted Fully Informed Particle Swarm Optimization variant. We evaluate these approaches on the Max-Cut problem for weighted 3-regular graphs~\cite{KARONSKI2004151, Commander2008}. Our study comprises three primary experimental settings: exact statevector simulations, shot based simulations incorporating measurement noise, and experiments on simulated quantum hardware. In addition, we perform a hyperparameter sweep and a study of swarm size to assess the sensitivity of the proposed methods. Performance is quantified using the metric $(1 - r)$, where $r$ denotes the approximation ratio, defined as the ratio of the obtained cut value to the optimal cut value.

\section{Preliminaries}

\subsection{QAOA and Swarm Optimization}

QAOA is naturally formulated through a Quadratic Unconstrained Binary Optimization representation of the underlying combinatorial problem \cite{Lewis2017}. In a weighted 3 regular graph, each vertex has degree three and edges are assigned real valued weights $w_{ij}$.

For a graph $G=(V,E)$, the weighted Max Cut problem in QUBO form is given by
\begin{equation}
\max_{\mathbf{x} \in {0,1}^n}
\sum_{(i,j)\in E}
w_{ij}
\left(x_i + x_j - 2x_i x_j\right).
\end{equation}

Introducing spin variables $z_i \in {-1,+1}$ via the transformation $z_i = 2x_i - 1$, the objective can be written as
\begin{equation}
C(\mathbf{z})
=
\frac{1}{2}
\sum_{(i,j)\in E}
w_{ij}
\left(1 - z_i z_j\right),
\end{equation}
which corresponds directly to the weighted cut value.

Promoting the classical spin variables to Pauli operators, $z_i \mapsto Z_i$, yields the cost Hamiltonian
\begin{equation}
H_C
=
\frac{1}{2}
\sum_{(i,j)\in E}
w_{ij}
\left(I - Z_i Z_j\right),
\end{equation}
which is diagonal in the computational basis and whose eigenvalues correspond to the classical objective values.

The QAOA variational state at circuit depth $p$ is constructed through alternating evolution under the cost and mixer Hamiltonians:
\begin{equation}
|\psi_p(\gamma,\beta)\rangle
=
\prod_{k=1}^{p}
e^{-i \beta_k H_M}
e^{-i \gamma_k H_C}
|+\rangle^{\otimes n},
\end{equation}
where the mixer Hamiltonian is defined as
\begin{equation}
H_M = \sum_{i\in V} X_i,
\end{equation}
with $X_i$ denoting Pauli $X$ operators and $|+\rangle^{\otimes n}$ the uniform superposition state.

The variational objective is defined as the expectation value of the cost Hamiltonian:
\begin{equation}
F_p(\gamma,\beta)
=
\langle
\psi_p(\gamma,\beta)
|
H_C
|
\psi_p(\gamma,\beta)
\rangle,
\label{eq:objective}
\end{equation}
and the associated optimization problem is
\begin{equation}
(\gamma^\star,\beta^\star)
=
\arg\max_{\gamma,\beta}
F_p(\gamma,\beta).
\label{eq:opt}
\end{equation}

To solve the optimization problem in Eq.~(\ref{eq:opt}), we employ swarm based optimization methods operating directly in the QAOA parameter space. Each particle $i$ represents a candidate parameter vector $\theta_i = (\gamma, \beta) \in \mathbb{R}^{2p}$, and its fitness is evaluated via the quantum objective $F_p(\theta_i)$. Maximizing this objective directly increases the approximation ratio,
\begin{equation}
r = \frac{F_p(\gamma, \beta)}{C^*},
\end{equation}
where $C^*$ denotes the optimal MaxCut value. Consequently, the performance metric $1 - r$ is minimized as the objective improves.



Let $\{\theta_i^{(t)}\}_{i=1}^{M}$ denote the swarm at iteration $t$, with corresponding fitness values $F_p(\theta_i^{(t)})$. Each particle maintains a personal best $\mathbf{p}_i^{(t)}$, while $\mathbf{g}^{(t)}$ denotes the global best solution. Different swarm optimization algorithms define distinct update dynamics. In the following, we describe the specific update rules for each method considered in this work.

\paragraph{Particle Swarm Optimization (PSO).}
Particle Swarm Optimization is a population-based stochastic optimization method in which a set of particles explore the search space by iteratively updating their velocities and positions based on both individual and collective experience \cite{Kennedy1995}. Each particle $i$ maintains its current position $\theta_i^{(t)}$ and velocity $v_i^{(t)}$ at iteration $t$. The update rules are given by
\begin{align}
v_i^{(t+1)} &=
w\,v_i^{(t)}
+ c_1\,r_{1,i}^{(t)} (p_i^{(t)} - \theta_i^{(t)})
+ c_2\,r_{2,i}^{(t)} (g^{(t)} - \theta_i^{(t)}), \\
\theta_i^{(t+1)} &=
\theta_i^{(t)} + v_i^{(t+1)},
\end{align}
where $p_i^{(t)}$ denotes the personal best position found so far by particle $i$, and $g^{(t)}$ represents the global best position discovered by the swarm up to iteration $t$. The superscript $(t)$ indicates the iteration index, while $(t+1)$ denotes the updated values at the next step. The parameter $w$ is the inertia weight controlling momentum, and $c_1$, $c_2$ are acceleration coefficients weighting the influence of personal and global information, respectively. The terms $r_{1,i}^{(t)}$ and $r_{2,i}^{(t)}$ are random variables sampled uniformly from $[0,1]$, introducing stochasticity into the search. 

\paragraph{Fully Informed Particle Swarm Optimization (FIPSO).}
Fully Informed Particle Swarm Optimization extends the standard PSO framework by allowing each particle to be influenced by multiple neighbors rather than only its personal and global best positions. This leads to a more distributed information sharing mechanism, which can improve exploration of the search space\cite{Mendes2004}. Each particle $i$ updates its velocity and position at iteration $t$ according to
\begin{align}
v_i^{(t+1)} &=
w\,v_i^{(t)}
+ \frac{c}{M} \sum_{j=1}^{M}
r_{ij}^{(t)} (p_j^{(t)} - \theta_i^{(t)}), \\
\theta_i^{(t+1)} &=
\theta_i^{(t)} + v_i^{(t+1)},
\end{align}
where $p_j^{(t)}$ denotes the best position found by neighbor $j$, and $M$ is the number of neighbors influencing particle $i$. The term $r_{ij}^{(t)}$ represents a random variable sampled uniformly from $[0,1]$ for each interaction between particles $i$ and $j$, introducing stochastic weighting of neighbor contributions. The superscript $(t)$ indicates the iteration index. The parameter $w$ controls inertia, while $c$ determines the overall strength of attraction toward neighboring best positions. 

\paragraph{Quantum Particle Swarm Optimization (QPSO).}
Quantum Particle Swarm Optimization reformulates the standard PSO dynamics using principles inspired by quantum mechanics, eliminating explicit velocity updates and instead modeling particle motion probabilistically. Each particle $i$ is attracted toward a stochastic local attractor that combines its personal best and the global best positions\cite{Sun2004ParticleSO}. The update rules are given by
\begin{align}
p_i^{*,(t)} &=
\phi_i^{(t)} p_i^{(t)}
+ (1 - \phi_i^{(t)}) g^{(t)}, \\
m^{(t)} &= \frac{1}{M} \sum_{j=1}^{M} p_j^{(t)}, \\
\theta_i^{(t+1)} &=
p_i^{*,(t)}
\pm \alpha^{(t)} \left| m^{(t)} - \theta_i^{(t)} \right|
\ln\left(\frac{1}{u_i^{(t)}}\right),
\end{align}
where $p_i^{(t)}$ and $g^{(t)}$ denote the personal and global best positions, respectively, and $\phi_i^{(t)} \in [0,1]$ is a random weighting factor that determines the contribution of each. The quantity $m^{(t)}$ represents the mean best position of all $M$ particles, providing a measure of the swarm’s central tendency. The variable $u_i^{(t)} \in (0,1)$ is a uniformly distributed random number, and $\alpha^{(t)}$ is a contraction expansion coefficient controlling the search scale. The superscript $(t)$ denotes the iteration index. The stochastic logarithmic term enables particles to explore the search space with enhanced global search capability.

\paragraph{Adam-assisted FIPSO (Adam-FIPSO).}
The Adam optimizer is directly incorporated into the Fully Informed Particle Swarm Optimization (FIPSO) framework, where its update rules are used to introduce first order momentum and adaptive scaling into the particle dynamics. This integration embeds Adam based gradient estimates into the velocity update mechanism, allowing particle updates to be guided by momentum terms and adaptively scaled adjustments derived from the optimizer.


\begin{align}
\Delta_i^{(t)} &=
w\,v_i^{(t)}
+ \frac{c}{M} \sum_{j=1}^{M}
r_{ij}^{(t)} \odot (p_j^{(t)} - \theta_i^{(t)}), \\
m_i^{(t)} &= \beta_1 m_i^{(t-1)} + (1 - \beta_1)\Delta_i^{(t)}, \\
v_i^{(t)} &= \beta_2 v_i^{(t-1)} + (1 - \beta_2)(\Delta_i^{(t)})^2, \\
\theta_i^{(t+1)} &=
\theta_i^{(t)}
+ \eta \frac{\hat{m}_i^{(t)}}{\sqrt{\hat{v}_i^{(t)}} + \epsilon}.
\end{align}

In this formulation, each particle update $\Delta_i^{(t)}$ is computed using the fully informed strategy, where information from all neighboring particles contributes to the direction of movement. The Adam inspired moment estimates $m_i^{(t)}$ and $v_i^{(t)}$ represent the first and second moments of the update signal, respectively, enabling smoother and more stable updates. Bias corrected estimates $\hat{m}_i^{(t)}$ and $\hat{v}_i^{(t)}$ are used to prevent initialization bias, while the adaptive learning rate term ensures that step sizes are scaled appropriately for each dimension.

\begin{figure*}[t]
    \centering
    \includegraphics[width=\textwidth]{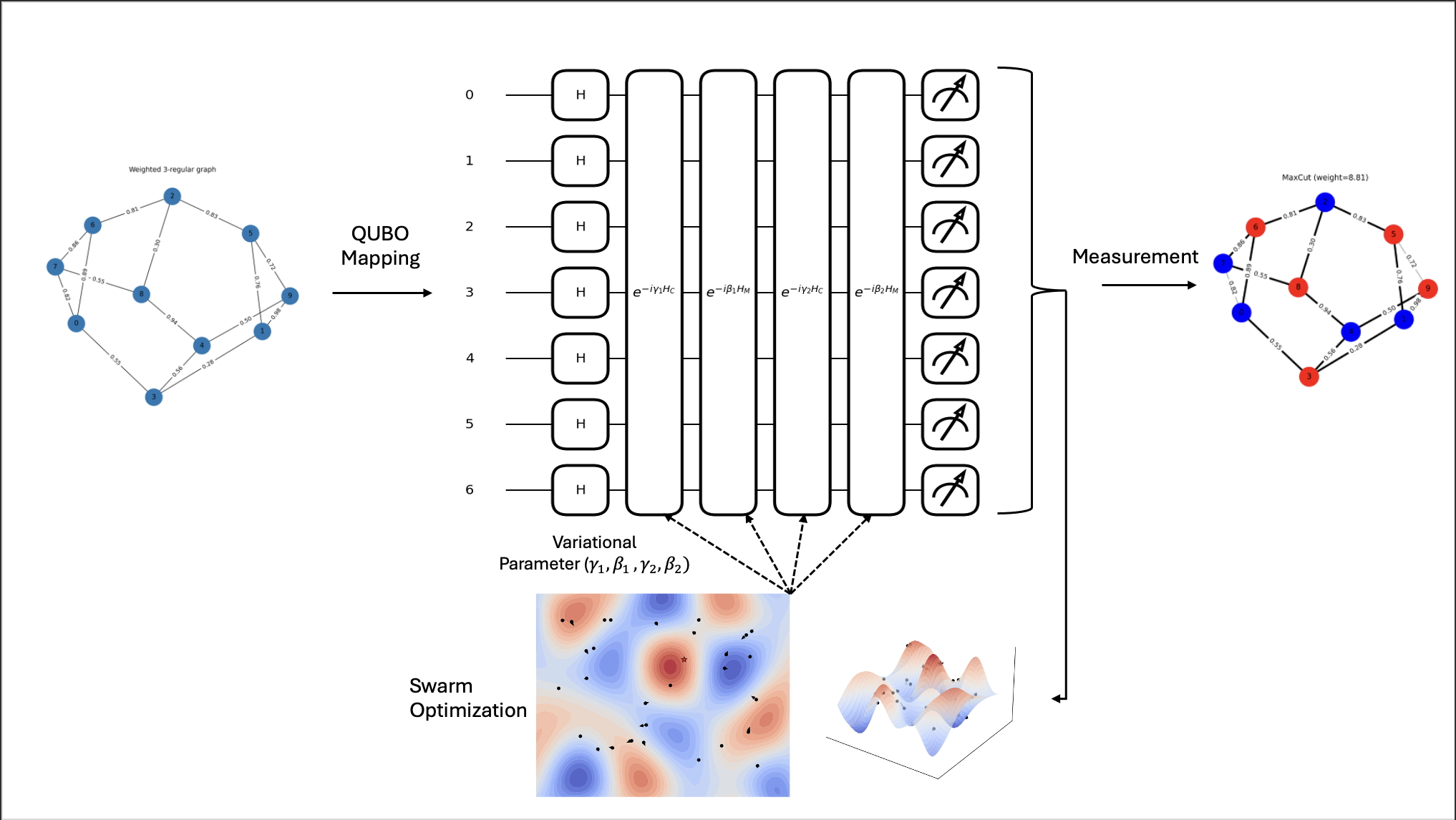}
    \caption{
    QAOA pipeline illustrating the workflow from a weighted 3-regular graph input through QUBO mapping and circuit construction, to optimization and final measurement. The QAOA circuit comprises parameterized cost and mixer layers, whose variational parameters are iteratively updated using swarm based optimization algorithms. At each iteration, the objective function is evaluated and used to guide the parameter updates, with the goal of minimizing the approximation gap. The final output corresponds to the Max-Cut solution.
    }
    \label{fig:qaoa_pipeline}
\end{figure*}


\subsection{Methodology}
Figure \ref{fig:qaoa_pipeline} illustrates the overall workflow of the swarm-based optimization framework applied to the QAOA circuit. A weighted 3 regular graph is first mapped to a QUBO formulation, which is subsequently encoded into the QAOA ansatz. The circuit comprises alternating cost and mixer layers, parameterized by $\gamma$ and $\beta$, respectively.


At each iteration, candidate parameter sets are generated by the swarm optimization algorithms and evaluated with respect to the objective function defined in Eq.~(\ref{eq:objective}). The resulting objective values guide the update of the swarm, and the refined parameters are fed back into the QAOA circuit. This iterative procedure is performed for a fixed budget of 800 iterations. The final optimized parameters are recorded, and the approximation gap is monitored throughout the optimization process. 

We consider five experimental settings: statevector simulation, shot-based simulation, execution on simulated (fake) hardware backends, a hyperparameter sweep, and a dedicated study of swarm optimization strategies. The next section presents the experimental results obtained under these settings, beginning with the statevector simulations.

\section{Results}
\subsection{Statevector Simulation} \label{sec:exactsim}
\begin{figure*}[h!]
    \centering
    \includegraphics[width=\textwidth]{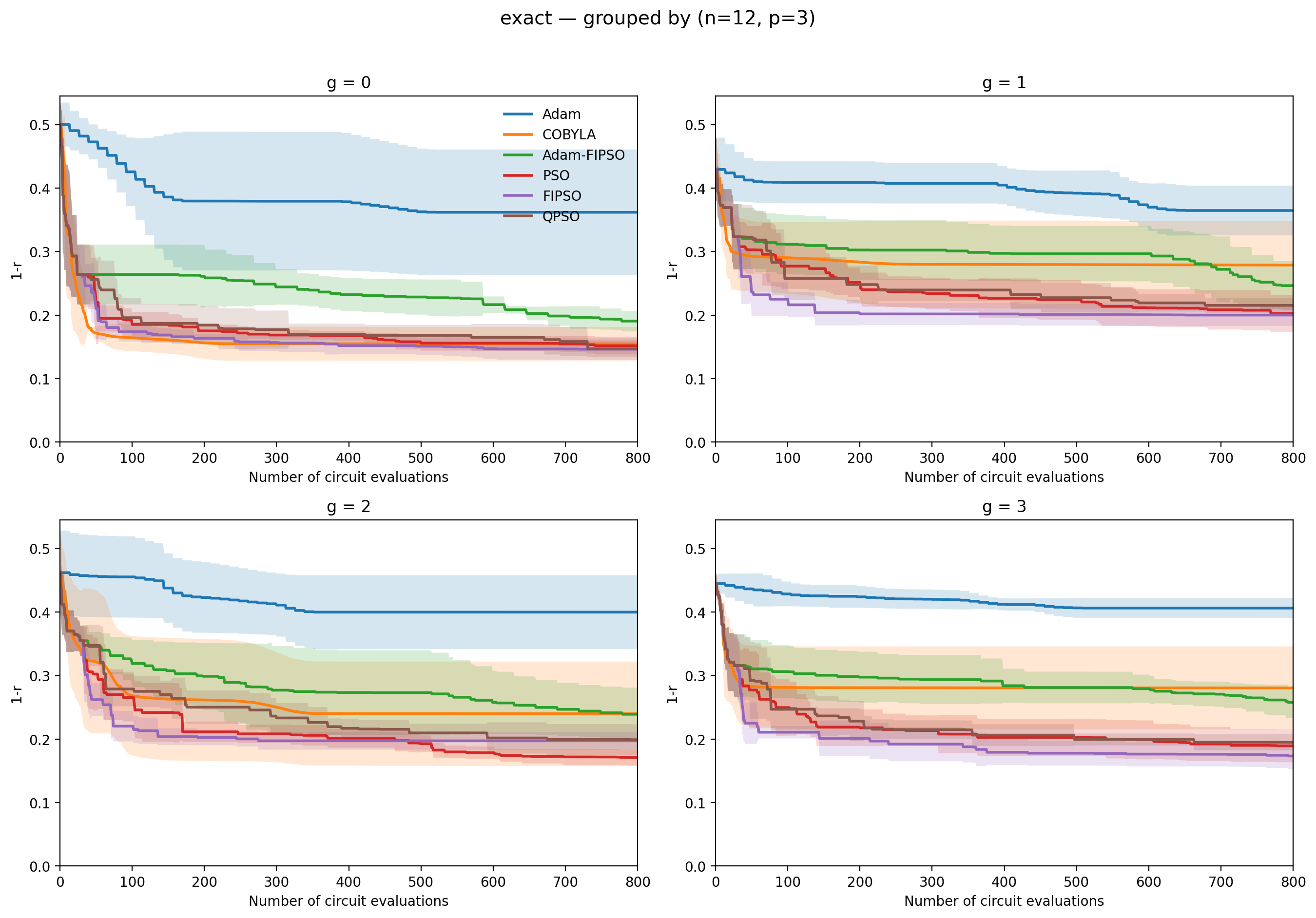}
    \caption{
    Statevector simulation results for graph size $n = 12$ at circuit depth $p = 3$. The plots show the convergence of different optimization algorithms in terms of the approximation error $(1 - r)$ as a function of the number of circuit evaluations. Each panel corresponds to a different problem instance (labeled by $g$). Solid lines represent the mean performance across runs, while the shaded regions indicate variability (e.g., standard deviation). 
    }
    \label{fig:grp_exact_123}
\end{figure*}

\begin{figure*}[h!]
    \centering
    \includegraphics[width=\textwidth]{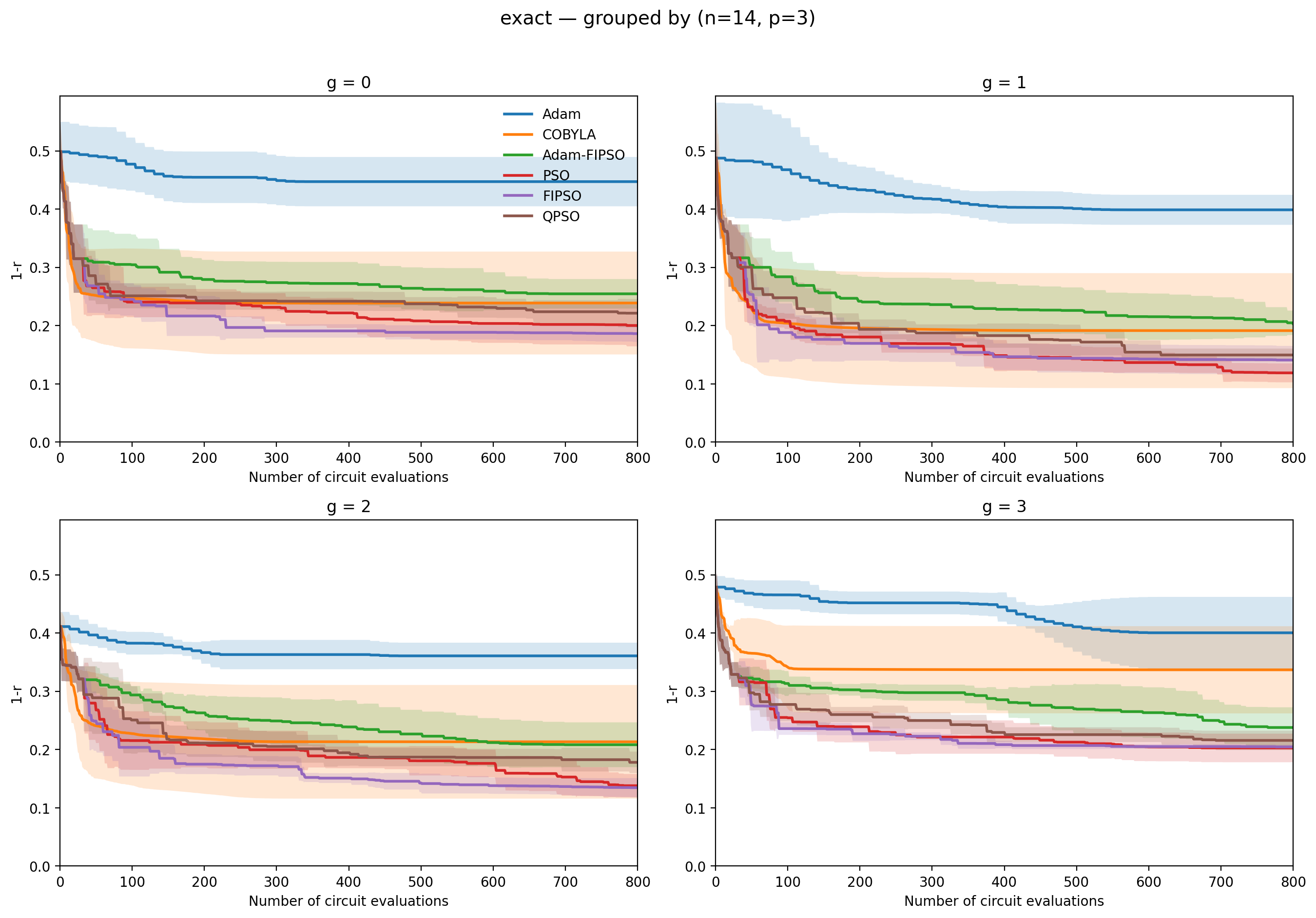}
    \caption{
    Statevector simulation results for graph size $n = 14$ at circuit depth $p = 3$. The plots show the convergence of different optimization algorithms in terms of the approximation error $(1 - r)$ as a function of the number of circuit evaluations. Each panel corresponds to a different problem instance (labeled by $g$). Solid lines represent the mean performance across runs, while the shaded regions indicate variability (e.g., standard deviation). 
    }
    \label{fig:grp_exact_143}
\end{figure*}

For the first experiment, we consider a noiseless environment and perform the optimization for graph sizes $n=10,12$ and $14$ and circuit depths $p=1,2,3,4$. The swarm based methods are benchmarked against Adam and COBYLA. Figures \ref{fig:grp_exact_123} and \ref{fig:grp_exact_143} show the exact simulation results across two system sizes (n=12 and n=14) at a circuit dept p=3. Across both graph sizes we observe qualitatively consistent optimization behaviour across all instances. Gradient based optimization (Adam) converges slowly and systematically plateaus at higher approximation error 1-r, indicating limited ability to escape suboptimal regions of the parameter landscape. In contrast, derivative free and population based methods exhibit rapid initial improvement and achieve lower final errors. Swarm based approaches (FIPSO and QPSO) consistently provide the best performance, suggesting improved exploration of the non-convex objective. Hybrid strategies (Adam–FIPSO) offer moderate gains over pure gradient-based methods but do not match the performance of fully population based optimizers. Increasing the system size from n=12 to n=14 does not cause a noticeable performance degradation: both convergence rates and final error levels remain comparable, and the relative ranking of optimizers is preserved in all graph instances.

\subsection{Shot based simulation}
In the shot based and Qiskit fake hardware settings, we additionally include SPSA as a noise resilient baseline. The graph sizes are kept consistent with the statevector simulations, i.e., $n = 10, 12, 14$, across circuit depths $p = 1, 2, 3, 4$. The number of measurement shots is varied across $50, 100, 1000,$ and $5000$.

\begin{figure*}[h!]
    \centering
    \includegraphics[width=\textwidth]{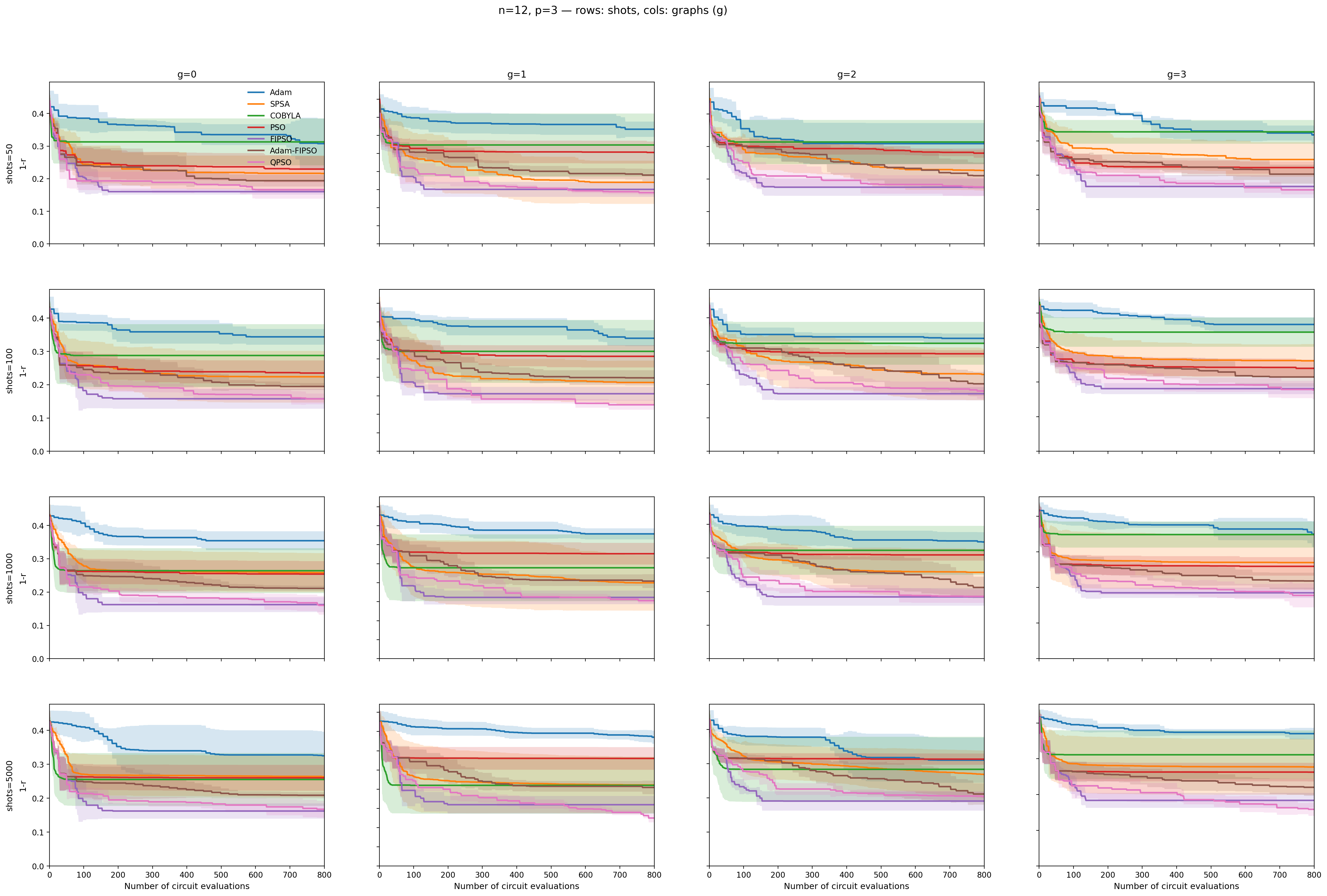}
    \caption{
    Shot-based simulation results for graph size $n = 12$ at circuit depth $p = 3$. The approximation error $(1 - r)$ is shown as a function of circuit evaluations. Columns correspond to different problem instances ($g = 0,1,2,3$), and rows to increasing shot budgets (50 to 5000). Solid lines denote mean performance, with shaded regions indicating variability. Increasing the number of shots reduces noise and improves convergence stability, with diminishing returns at higher shot counts. Swarm-based methods (FIPSO and QPSO) consistently achieve superior performance across all settings.
    }
    \label{fig:grp_shot_123}
\end{figure*}

\begin{figure*}[h!]
    \centering
    \includegraphics[width=\textwidth]{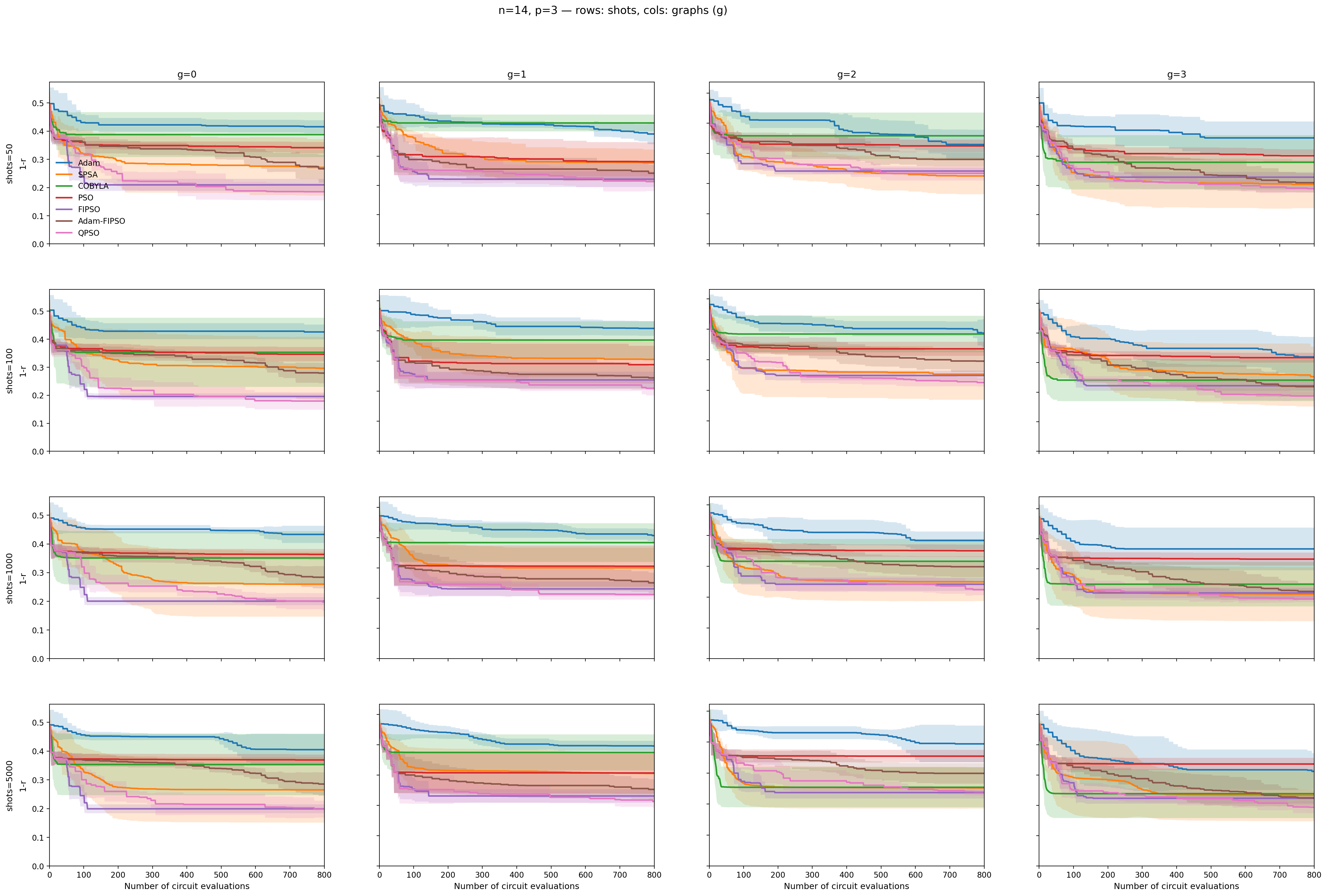}
    \caption{
    Shot-based simulation results for graph size $n = 14$ at circuit depth $p = 3$. The approximation error $(1 - r)$ is shown as a function of circuit evaluations. Columns correspond to different problem instances ($g = 0,1,2,3$), and rows to increasing shot budgets (50 to 5000). Solid lines denote mean performance, with shaded regions indicating variability. Increasing the number of shots reduces noise and improves convergence stability, with diminishing returns at higher shot counts. Swarm-based methods (FIPSO and QPSO) consistently achieve superior performance across all settings.
    }
    \label{fig:grp_shot_143}
\end{figure*}

Figures~\ref{fig:grp_shot_123} and~\ref{fig:grp_shot_143} present the optimization performance across different shot budgets for weighted 3-regular MaxCut instances. Increasing the number of shots systematically reduces statistical fluctuations in the objective evaluation, resulting in smoother optimization trajectories and improved convergence stability. The most pronounced improvements are observed when moving from low (50) to intermediate (100--1000) shot regimes. Beyond this range, performance gains become progressively marginal, indicating diminishing returns at higher shot counts.

The effect of shot noise differs across optimizers. Gradient based methods such as Adam are highly sensitive to noise. Measurement noise affects gradient estimates, which leads to unstable updates, slower convergence, and early saturation at higher approximation error. SPSA is more robust to noise due to its stochastic nature, but it still struggles to consistently reach low error solutions. In contrast, derivative free methods such as COBYLA and population based methods such as PSO show faster initial improvement and are less affected by shot noise. These methods maintain more stable convergence across different shot budgets. Swarm based approaches including FIPSO and QPSO achieve the best performance, reaching lower final errors with reduced variation across runs. The hybrid Adam FIPSO method improves over standard gradient based optimization, but still does not match the performance of purely population based approaches.


\subsection{Qiskit Fake Hardware Simulation}

The experiments are conducted on a simulated 7-qubit superconducting quantum device based on the Fake Nairobi backend\footnote{\url{https://quantum.cloud.ibm.com/docs/en/api/qiskit-ibm-runtime/fake-provider-fake-nairobi-v2}}, which emulates realistic hardware characteristics. The native gate set comprises single-qubit operations (\textit{id}, \textit{rz}, \textit{sx}, \textit{x}) and two-qubit entangling gates (\textit{cx}), together with a non-fully connected coupling map that constrains qubit interactions. 

The device exhibits average relaxation and dephasing times of $T_1 \approx 96.15\,\mu s$ and $T_2 \approx 84.24\,\mu s$, respectively, indicating moderate coherence suitable for near-term quantum circuits. Gate fidelity analysis shows low average single-qubit error rates on the order of $3.15 \times 10^{-4}$, while two-qubit error rates are higher, averaging $8.77 \times 10^{-3}$. Readout errors vary across qubits, ranging approximately from $1.8\%$ to $5.8\%$, thereby providing a realistic noisy intermediate-scale quantum (NISQ) environment for evaluating algorithmic performance under hardware constraints.

\begin{figure*}[h!]
    \centering
    \includegraphics[scale=0.6]{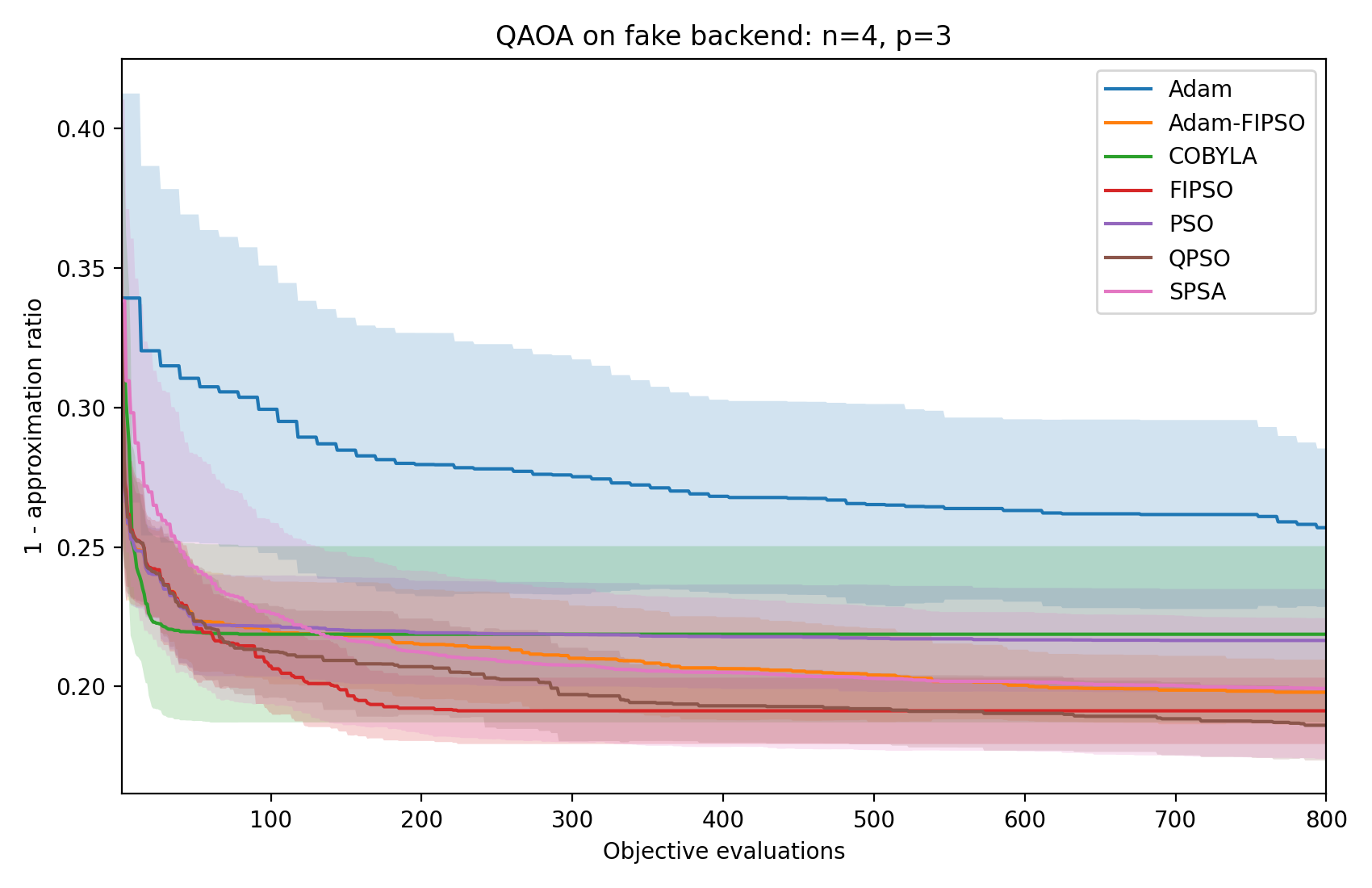}
    \caption{
    Fake hardware simulation results for graph size $n = 4$ at circuit depth $p = 3$. The approximation error $(1 - r)$ is shown as a function of the number of objective evaluations. Solid lines represent mean performance across runs, with shaded regions indicating variability. Swarm-based methods (FIPSO and QPSO) achieve faster convergence and lower final errors compared to gradient based (Adam) and derivative free (COBYLA) optimizers, while SPSA provides a competitive noise resilient baseline under realistic hardware noise.
    }
    \label{fig:fh_43}
\end{figure*}

\begin{figure*}[h!]
    \centering
    \includegraphics[scale=0.6]{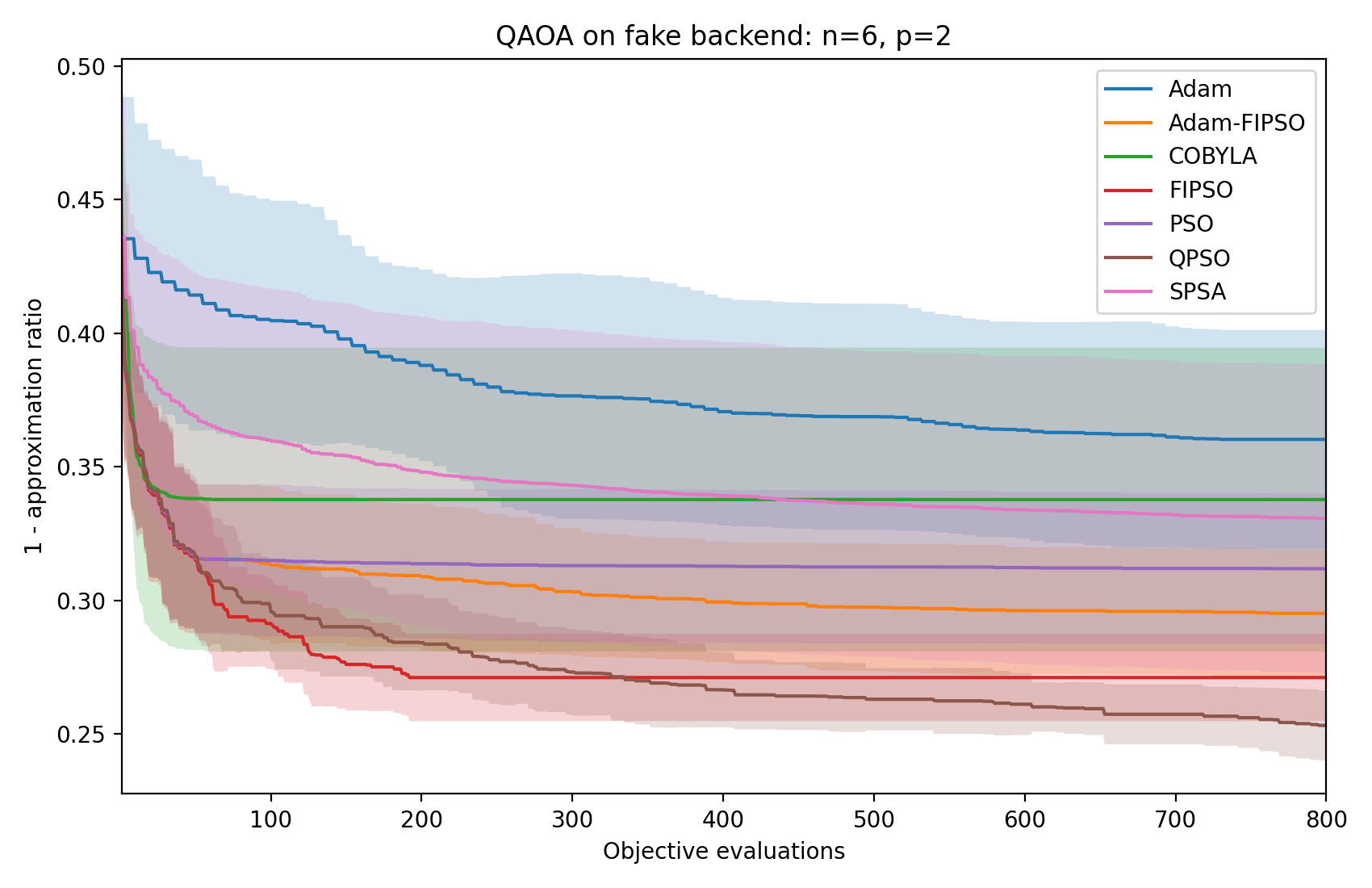}
    \caption{
    Fake hardware simulation results for graph size $n = 6$ at circuit depth $p = 2$. The approximation error $(1 - r)$ is shown as a function of the number of objective evaluations. Solid lines represent mean performance across runs, with shaded regions indicating variability. Swarm-based methods (FIPSO and QPSO) achieve faster convergence and lower final errors compared to gradient based (Adam) and derivative free (COBYLA) optimizers.
    }
    \label{fig:fh_62}
\end{figure*}

Due to the computational cost of simulating QAOA in a noisy setting, we restrict our analysis to smaller graph sizes $n = 4$ and $n = 6$ across circuit depths $p = 1, 2, 3$. Figures~\ref{fig:fh_43} and~\ref{fig:fh_62} show the convergence of the optimizers for graph sizes $4$ and $6$, respectively. The experiment with $n = 4$ uses a circuit depth of $p = 3$, while the experiment with $n = 6$ uses $p = 2$.

In the $n=4, p=3$ case, all optimizers exhibit rapid initial improvement, followed by a gradual convergence phase. Among the methods, FIPSO and QPSO consistently achieve the lowest final values, indicating superior solution quality, with FIPSO converging faster in the early stages and stabilizing around $\sim 0.19$. The Adam-FIPSO hybrid demonstrates competitive performance, outperforming standard PSO, COBYLA, and SPSA, and achieving a lower asymptotic value than Adam alone. Adam shows relatively slower convergence and higher final error, suggesting that purely gradient-based updates may be less effective in this landscape. COBYLA converges quickly but plateaus early, indicating limited exploration capability.

For the $n=6, p=2$ setting, which represents a slightly more challenging optimization problem, the differences between optimizers become more pronounced. FIPSO again achieves strong performance, reaching a low final value around $\sim 0.27$, while QPSO continues to improve steadily over the full evaluation budget, ultimately approaching similar performance. Adam-FIPSO demonstrates improved convergence compared to standalone Adam and PSO, maintaining a consistent downward trend and achieving a favorable trade-off between convergence speed and final accuracy. SPSA shows stable but slower improvement, whereas COBYLA quickly reaches a plateau at a relatively high error level. Adam again exhibits slower and less effective convergence, highlighting limitations in noisy, non-convex quantum optimization settings.

\subsection{Hyperparameter Sweep}
The hyperparameter sweep experiments are performed for graph sizes $n = 10, 12$ and circuit depths $p = 3, 4$ to study the effect of varying hyperparameters across problem instances and circuit depth. For PSO, we vary the inertia weight $w$ and the coefficients $c_1$ and $c_2$, which control the attraction toward personal and global best positions. In FIPSO, the dynamics are governed by the inertia weight $w$ and a scaling parameter $c$ that determines the influence of other particles. Adam-FIPSO extends this update by introducing a learning rate parameter that controls the step size through adaptive moment estimation. QPSO uses a contraction coefficient $\alpha$, which is annealed from $\alpha_{\mathrm{start}}$ to $\alpha_{\mathrm{end}}$ over the course of optimisation. All methods use a swarm size of 30 and are evaluated under a fixed budget of 800 circuit evaluations.

\begin{figure*}[h!]
    \centering
    \includegraphics[width=\textwidth]{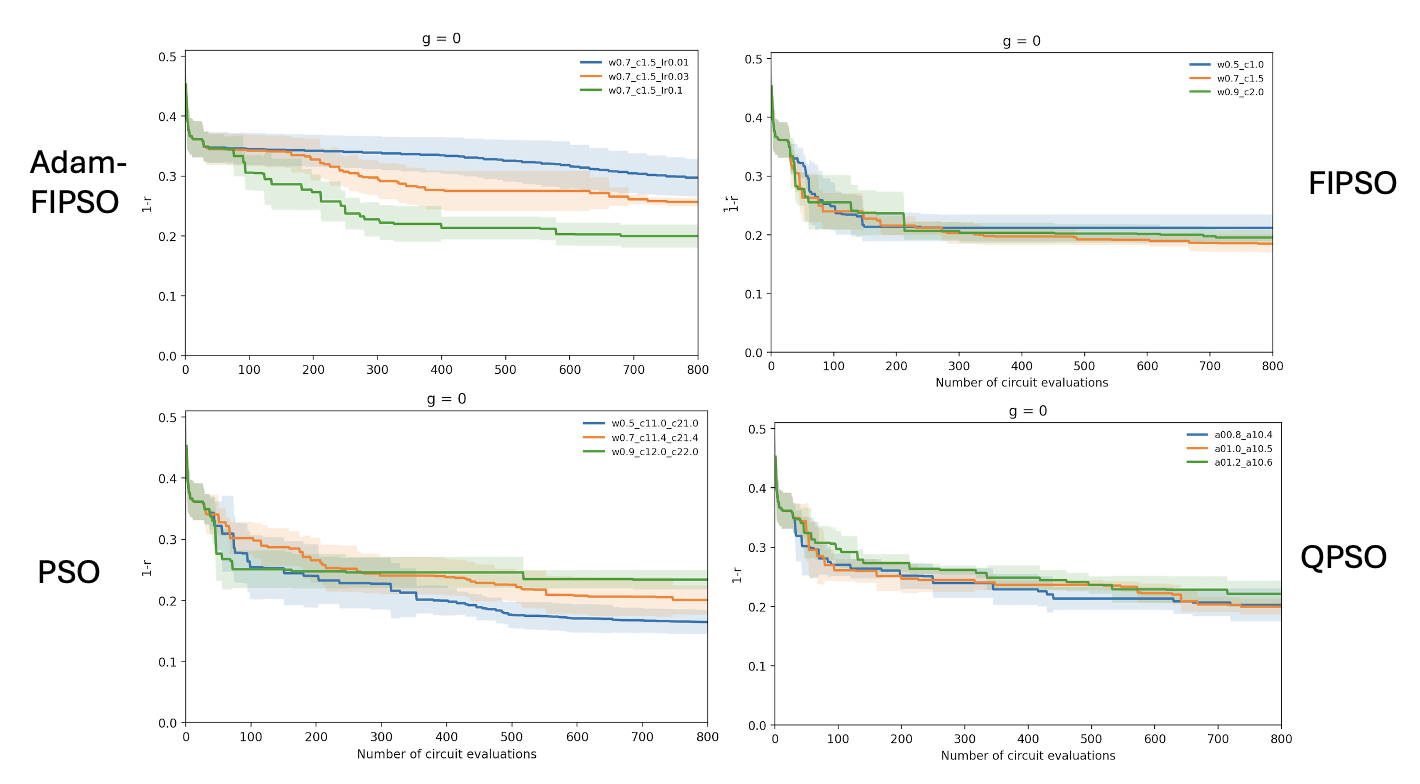}
    \caption{
     Hyperparameter sweep results for graph size $n = 10$ at circuit depth $p = 4$. The approximation error $(1 - r)$ is shown as a function of circuit evaluations for different optimizer configurations. Each panel corresponds to a different method (Adam-FIPSO, FIPSO, PSO, and QPSO), with curves representing distinct hyperparameter settings. Solid lines denote mean performance, with shaded regions indicating variability. Performance is sensitive to hyperparameter choices, with well tuned swarm based methods achieving faster convergence and lower final errors.
    }
    \label{fig:hs}
\end{figure*}

Figure \ref{fig:hs} compares the statevector simulation convergence behaviour of Adam-FIPSO, FIPSO, PSO, and QPSO under different hyperparameter configurations for the graph size n=10 and circuit depth p=4. Clear differences in sensitivity to hyperparameter choice are observed across the optimisers. Adam-FIPSO exhibits the strongest dependence, with certain configurations converging significantly faster and to lower values than others, indicating that performance is highly contingent on careful tuning. In contrast, FIPSO shows rapid initial improvement followed by early saturation, with trajectories that remain relatively close across configurations, suggesting greater robustness. PSO displays intermediate behaviour, with noticeable but less pronounced separation between configurations and a more gradual convergence profile. QPSO shows the least variation, with closely clustered curves and similar convergence dynamics across all settings, indicating stable but less tunable performance. 

\subsection{Swarm Study}
The final experiment investigates the effect of swarm size on the convergence behaviour of the optimisers. We consider swarm sizes of 10, 20, 50, 70, and 100, evaluated on graph sizes $n = 10$ and $n = 12$ across circuit depths $p = 3$ and $p = 4$. The hyperparameters for all swarm-based optimisers are kept fixed to isolate the impact of swarm size. For PSO, the inertia weight is set to $w = 0.7$, while the cognitive and social coefficients are set to $c_1 = 1.4$ and $c_2 = 1.4$. For FIPSO, the inertia weight remains $w = 0.7$ and the scaling parameter is set to $c = 1.5$. Adam-FIPSO uses the same values of $w$ and $c$, with a learning rate of $0.03$. For QPSO, the contraction coefficient is annealed from $\alpha_{\mathrm{start}} = 1.0$ to $\alpha_{\mathrm{end}} = 0.5$. All optimisers are evaluated under a fixed budget of 800 circuit evaluations.

\begin{figure*}[t]
    \centering
    \includegraphics[width=\textwidth]{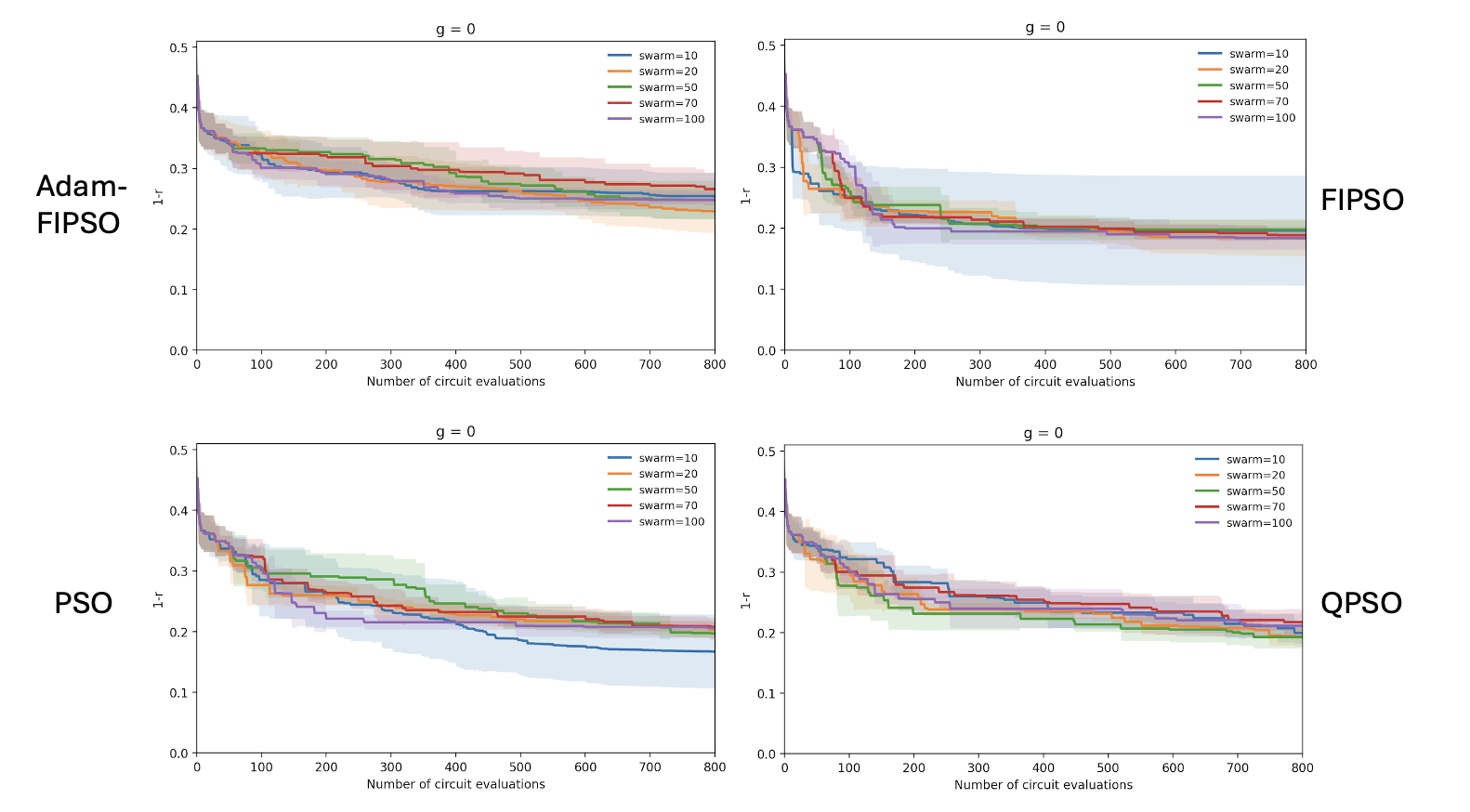}
    \caption{
     Swarm size study for graph size $n = 10$ at circuit depth $p = 4$. The approximation error $(1 - r)$ is shown as a function of circuit evaluations for different swarm sizes. Each panel corresponds to a different optimization method (Adam-FIPSO, FIPSO, PSO, and QPSO). Solid lines denote mean performance across runs, with shaded regions indicating variability. Increasing swarm size generally improves convergence stability, with diminishing returns beyond moderate swarm sizes.
    }
    \label{fig:ss}
\end{figure*}

Figure \ref{fig:ss} shows the statevector simulation results of varying swarm size for Adam-FIPSO, FIPSO, PSO, and QPSO for the instance n=10 and p=4. For Adam-FIPSO, all swarm sizes exhibit similar convergence trajectories and comparable final performance, with overlapping variance bands, indicating both low sensitivity to swarm size and consistent stability across runs. In FIPSO, larger swarm sizes lead to faster initial convergence, but this comes at the cost of increased variance in the early stages, particularly for the largest swarm, after which all configurations stabilise to similar performance with reduced variability. PSO shows a stronger dependence on swarm size, where smaller swarms achieve better final performance and exhibit tighter variance bands, while larger swarms converge more slowly and display increased variability, suggesting less stable behaviour under a fixed evaluation budget. QPSO demonstrates relatively uniform behaviour across all swarm sizes, with closely clustered trajectories and similar variance throughout, indicating robustness both in terms of convergence and stability.

\section{Future Work}
Future work will focus on extending swarm based optimization methods to a broader class of variational quantum algorithms beyond the MaxCut problem, including applications in combinatorial optimization and quantum chemistry. Such studies will help determine whether the observed performance advantages generalize across different problem structures and Hamiltonians.

An important next step is validation on actual quantum hardware. While this work incorporates realistic noise through hardware calibrated simulation backends, real devices introduce additional challenges such as calibration drift, temporal noise fluctuations, and hardware instability. Evaluating swarm based methods under these conditions will be critical for assessing their practical robustness.

Scalability is another key direction. As system size and circuit depth increase, the dimensionality of the parameter space grows, potentially affecting convergence behavior and computational cost. Future work should investigate how swarm size and update dynamics scale in high dimensional regimes, and whether hybrid approaches can improve efficiency.

Finally, further analysis of optimization dynamics in variational quantum landscapes may provide deeper insight into the observed performance of population based methods. Such studies could inform the design of more effective and adaptive optimization strategies for near-term quantum computing applications

\section{Conclusion}
In this work we investigated swarm based optimization methods for parameter initialization and optimization in the Quantum Approximate Optimization Algorithm. A comprehensive set of experiments was conducted using exact statevector simulations shot based evaluations as well as noise aware hardware emulation. The results show that population based methods such as Particle Swarm Optimization, Fully Informed Particle Swarm Optimization and Quantum Particle Swarm Optimization consistently outperform widely used optimizers such as Adam COBYLA and SPSA.

Swarm based approaches achieve lower approximation gap and converge faster. Performance remains robust under noise as well as finite sampling conditions. Fully Informed Particle Swarm Optimization and Quantum Particle Swarm Optimization show stable behaviour across problem sizes circuit depths as well as noise regimes. The results indicate that population based exploration is well suited to non convex optimization landscapes in variational quantum algorithms. The hybrid Adam FIPSO method improves upon gradient based methods but does not match fully population based approaches.

This study focuses on the MaxCut problem on regular graphs. The observed advantages are expected to extend to other combinatorial optimization problems as well as variational quantum algorithms. The Adam FIPSO method is also highly sensitive to hyperparameters with performance varying significantly across different configurations in contrast to the more stable behaviour of other optimizers.

Overall this work establishes swarm based optimization as a practical and effective approach for improving QAOA performance in near term quantum computing. 

\newpage
\bibliographystyle{IEEEtran}
\bibliography{bibliography}

\EOD

\end{document}